\documentclass[
    ,final            
  ]
  {aipproc}

\layoutstyle{8x11double}

\begin{document}

\title{The Pulsed Spectra of Two Extraordinary Pulsars}

\author{Mallory Roberts}{
  address={Dept. of Physics, McGill University, 3600 University St. Montr\'eal,
QC. H3A 2T8, Canada},
  altaddress={Dept. of Physics and Center for Space Research, Massachusetts Institute
of Technology, Cambridge, MA. 02139 }
}

\author{Scott Ransom}{
  address={Dept. of Physics, McGill University, 3600 University St. Montr\'eal,
QC. H3A 2T8, Canada},
  altaddress={Dept. of Physics and Center for Space Research, Massachusetts Institute
of Technology, Cambridge, MA. 02139 }
}

\author{Fotis Gavriil}{
  address={Dept. of Physics, McGill University, 3600 University St. Montr\'eal,
QC. H3A 2T8, Canada},
}

\author{Vicky Kaspi}{
  address={Dept. of Physics, McGill University, 3600 University St. Montr\'eal,
QC. H3A 2T8, Canada},
}

\author{Pete Woods}{
  address={Universities Space Research Association}
  ,altaddress={National Space Science and Technology Center, 320 Sparkman Drive, Huntsville, AL 35805}
}

\author{Alaa Ibrahim}{
   address={NASA Goddard Space Flight Center, Greenbelt, MD 20771}
  ,altaddress={Department of Physics, George Washington University, Washington, DC 20006}
}

\author{Craig Markwardt}{
  address={NASA Goddard Space Flight Center, Greenbelt, MD 20771}
}

\author{Jean Swank}{
  address={NASA Goddard Space Flight Center, Greenbelt, MD 20771}
}

\begin{abstract}

We report on X-ray monitoring of two isolated pulsars within the same 
RXTE field of view. PSR J1811$-$1925 in the young supernova remnant
G11.2$-$0.3 has a nearly sinusoidal pulse profile with a hard
pulsed spectrum (photon index $\Gamma\sim 1.2$). 
The pulsar is a highly efficient ($\sim 1\%$ of spin-down energy) 
emitter of 2-50 keV pulsed X-rays despite having a fairly typical
$B\sim 2\times 10^{12}$~G magnetic field. PSR J1809$-$1943/XTE J1810$-$197
is a newly discovered slow ($P=5.54$~s), apparently isolated X-ray pulsar 
which increased in flux by a factor of 100 in 2003 January. Nine months
of monitoring observations have shown a decrease in pulsed flux of $\sim 30\%$ 
without a significant change in its apparently thermal spectrum ($kT 
\sim 0.7$~keV) or pulse profile.
During this time, the spin-down torque has 
fluctuated by a factor of $\sim 2$. Both the torque and the flux have remained
steady for the last 3 months, at levels consistent with a magnetar 
interpretation.

\end{abstract}

\maketitle


\section{Introduction to the G11.2$-$0.3 Field}

The region near the supernova remnant G11.2$-$0.3 is incredibly rich. 
Within a $\sim 1\,{\rm deg}^2$ are at least 5 supernova remnants \citep{bdl+03}
and several young pulsars. $\sim 1.2^{\circ}$ away is 
the Soft Gamma-Ray Repeater
SGR 1806$-$20 and not much further away is the possible SGR 1801$-$23. However, 
this region does not appear to be along a tangent line of a spiral arm or
especially abundant in star forming regions for an inner region of the 
Galactic plane. The region is within the area covered by the RXTE Galactic
bulge montoring program \citep{sm01}. This in combination with the facts that 
SNR G11.2$-$0.3 is one of only 6-8 remnants associated with historical events
\citep{cs77a} and SGR 1806$-$20 is one of only 4 confirmed SGRs has caused
this small region to be unusually well observed. Even so, it is quite
surprising that not only does the pulsar in SNR G11.2$-$0.3 have unusual
X-ray properties, but that while monitoring it with RXTE a transient 
Anomalous X-ray Pulsar appeared within the PCA field of view.
Here we report on the pulsed spectra of these two extraordinary pulsars. 

\section{PSR J1811$-$1925, the pulsar in SNR G11.2$-$0.3}

SNR G11.2$-$0.3 is remarkable in several ways.
It has a circular shell which is
bright in both radio and X-rays with a central pulsar wind nebula
\citep[see][for details of the nebular emission]{rtk+03,trk02}. 
Measurements of the expansion of the radio shell have established its age 
to be $\tau\sim 2,000$~yr \citep{tr03}, 
which is consistent with being born in the historical event of 386 A.D.
The PWN is powered
by one of the brightest X-ray emitting pulsars, PSR J1811$-$1925, 
whose characteristic age is  $\tau_c\sim 24,000$~yr \citep{ttd+99}. 
The discrepancy with the true age of the SNR implies it was born
spinning near its current spin period $P\sim 65$~ms \citep{krv+01}.
The pulsar's X-ray spectrum
is very hard ($\Gamma\sim 1$) and  has broad, nearly sinusoidal pulse profile reminiscent of the
profiles of PSR B1509$-$58 \citep{rjm+98} and the pulsar in the distant young remnant 
Kes75 \citep{gvbt00}, which are also very young 
spin-powered pulsars with very high hard X-ray efficiencies.

\begin{figure}
\label{g11prof}
  \includegraphics[height=.3\textheight, angle=270]{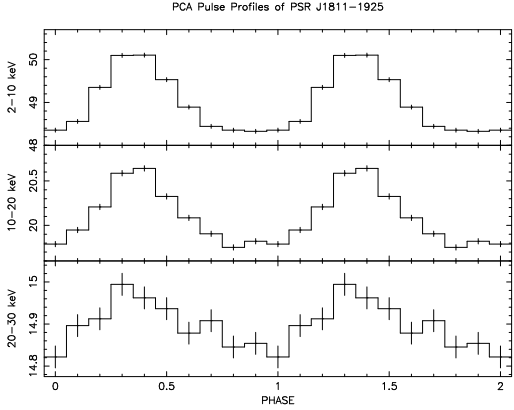}
  \caption{RXTE PCA pulse profiles of PSR J1811$-$1925}
\end{figure}

\begin{figure}
\label{g11profh}
  \includegraphics[height=.3\textheight, angle=270]{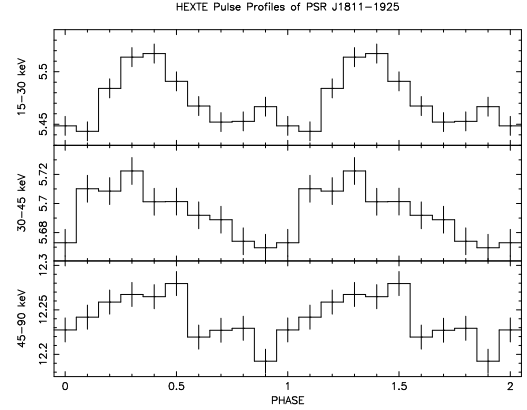}
  \caption{RXTE HEXTE pulse profiles of PSR J1811$-$1925}
\end{figure}

We have been monitoring PSR J1811$-$1925 with RXTE on a monthly basis since March 2002, 
with the hope of constraining the pulsar braking index to strengthen the case for a slow initial
period. The timing properties will be reported elsewhere. Here, we present the pulse profile
from the combined data set as a function of energy which is seen in both the PCA and HEXTE data (Fig.~\ref{g11prof}, \ref{g11profh}), and measurements of 
the hard X-ray pulsed spectrum. The pulsed emission is very hard, the spectrum 
being well fit in the 2.5-30 keV band by a single power law with
a photon index of $\Gamma = 1.16$ (0.97 - 1.36, 90\% confidence) 
with no sign of a cutoff in either the PCA or HEXTE data. The pulsations
are clearly seen above 45 keV in the HEXTE data, remaining
nearly sinusoidal. At the best determined 
distance of 5 kpc, assuming beaming into $\pi$ steradians implies a remarkable 
$\sim 1\%$ of the spin-down power 
is being emitted as pulsed 1-50 keV X-rays. The pulsed properties
of PSR J1811$-$1925 are very similar to those of the young ($\tau\sim 1700$~yr)
150~ms pulsar PSR B1509$-$58 \citep{rjm+98}. Both have broad profiles with no
apparent dependence on energy,  similar
spectra up to 50 keV, and similar X-ray efficiencies. The spectrum 
of PSR B1509$-$58 shows curvature which becomes apparent only 
above 50 keV \citep{cmm+01}, steepening to a power law of $\Gamma \sim 1.7$
before cutting off at $\sim 10$~MeV. 
The high X-ray efficiency of PSR J1811$-$1925
requires its emission to have curvature, a spectral break or cut-off below a 
few MeV, similar to PSR B1509$-$58 and leaves little power for GeV emission. 
However, unlike
PSR B1509$-$58 which has $B\sim 1.5\times 10^{13}$~G and about three
times the spin-down energy, or PSR J1846$-$0248 in Kes 75 which has an even higher
magnetic field, PSR J1811$-$1925 has a typical magnetic field of $\sim 2\times
10^{12}$~G. This suggests that unusually efficient magnetospheric pulsed 
X-ray emission results from a special geometry and not from a high magnetic 
field or spin-down energy, which may have important implications for
models of magnetospheric emission. Note as well that while PSR B1509$-$58
is a radio pulsar, neither PSR J1811$-$1925 or PSR J1846$-$0248 have
yet been seen in radio despite targetted searches \citep{ckm+01}. 


\section{A New AXP? XTE J1810$-$197 / PSR J1809$-$1943}

\begin{figure}
\label{1810_prof}
  \includegraphics[height=.3\textheight, angle=270]{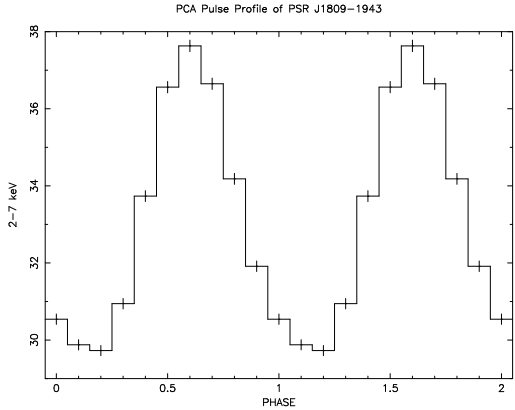}
  \caption{RXTE PCA 2-7 keV pulse profile of PSR J1809$-$1943}
\end{figure}

In July 2003, a TOO observation of the soft $\gamma$-ray repeater 
SGR 1806$-$20
revealed the previously undetected 5.54~s X-ray pulsar XTE J1810$-$197/
PSR J1809$-$1943 
\citep{mis03} with a sinusoidal profile 
(Fig.\ref{1810_prof}). Analysis of our 
G11.2$-$0.3 monitoring data showed the new pulsar first became detectable 
by RXTE in Jan. 2003. In combination
with data from the PCA monitoring program of the Galactic bulge region 
\citep{sm01}, 
a phase connected timing solution has been obtained which shows large torque 
variations but an
overall spin-down consistent with the new source being an anomalous X-ray 
pulsar \cite{ims+03}. 
Subsequent observations
with Chandra and XMM-Newton localized the source and measured its soft 
spectrum, which is
best fit by a two component blackbody ($kT\sim 0.3$~keV and $kT\sim 0.7$~keV). 
Archival ROSAT and ASCA observations detected the source at a flux level about 
a factor of 100 less than the outburst flux, with a temperature similar to 
the lower blackbody temperature observed by XMM-Newton
\citep{ghbb03}. 

\begin{figure}
\label{fluxhist}
  \includegraphics[height=.3\textheight, angle=270]{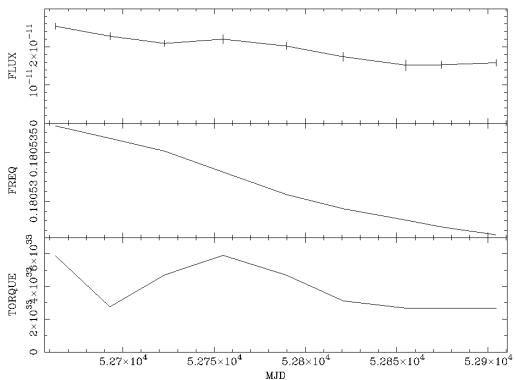}
  \caption{TOP: Flux history (in erg ${\rm cm}^{-2} {\rm s}^{-1}$) of
PSR J1809$-$1943. MIDDLE: 
Frequency (in ${\rm s}^{-1}$) evolution of PSR J1809$-$1943. 
BOTTOM: Spin-down torque (in erg ${\rm s}^{-1}$) evolution of PSR J1809$-$1943.}
\end{figure}

In Fig.\ref{fluxhist} we present the pulsed flux history of the transient X-ray pulsar PSR J1809$-$1943 as observed in the
G11.2$-$0.3 monitoring data covering the period from Jan. 2003 to Sep. 2003.  
All of the PCA observations are well fit by a single blackbody with 
$kT=0.7$~keV in the 2.5-8 keV range. There is no evidence for a change in the 
spectrum other than a drop of about 30\% in the overall flux. 
During this time, there were spin-down torque fluctuations of more than a 
factor of 2. For the last three observations, spanning about 90 days, both 
the flux and the torque have remained essentially constant. 
If the current torque is approximately that of dipole braking, then
the implied surface magnetic field is $B\sim 3\times 10^{14}$~G, on  
the low end of claimed magnetars. 
The absorption column towards the source is only 
$nH\sim 0.6 \times 10^{21}\,{\rm cm}^{-2}$, about one third that of G11.2$-$0.3,
suggesting it might be significantly closer than the latter's distance
of 5 kpc. This also makes this source a good target for infrared
studies. Even at a distance
of only 1 kpc, the X-ray luminosity is $\sim 5\times 10^{34}\,{\rm erg}^{-1}$,
significantly larger than the spin-down torque. Thus, it meets the basic
criteria for being an Anomalous X-ray pulsar. 

\begin{theacknowledgments}
This research was partially supported by NSERC Discovery and 
Steacie grants, the Canada Research Chairs, NATEQ, CIAR and NASA.
\end{theacknowledgments}


\bibliographystyle{aipproc}   

\bibliography{journals1,newrefs,psrrefs}

\begin{thebibliography}{15}
\expandafter\ifx\csname natexlab\endcsname\relax\def\natexlab#1{#1}\fi
\providecommand{\enquote}[1]{``#1''}
\expandafter\ifx\csname url\endcsname\relax
  \def\url#1{\texttt{#1}}\fi
\expandafter\ifx\csname urlprefix\endcsname\relax\def\urlprefix{URL }\fi

\bibitem[{Brogan} et~al.(2003)]{bdl+03}
{Brogan}, C.~L., {Devine}, K.~E., {Lazio}, T.~J., {Kassim}, N.~E., {Tam},
  C.~R., {Brisken}, W.~F., {Dyer}, K.~K., and {Roberts}, M.~S.~E., \emph{ArXiv
  Astrophysics e-prints} (2003).

\bibitem[{Swank} and {Markwardt}(2001)]{sm01}
{Swank}, J., and {Markwardt}, C., \enquote{{Populations of Transient Galactic
  Bulge X-ray Sources},} in \emph{ASP Conf. Ser. 251: New Century of X-ray
  Astronomy}, 2001, pp. 94--+.

\bibitem[Clark and Stephenson(1977)]{cs77a}
Clark, D.~H., and Stephenson, F.~R., \emph{The Historical Supernovae},
  Pergamon, Oxford, 1977.

\bibitem[Roberts et~al.(2003)]{rtk+03}
Roberts, M. S.~E., Tam, C.~R., Kaspi, V.~M., Lyutikov, M., Vasisht, G.,
  Pivovaroff, M., Gotthelf, E.~V., and Kawai, N., \emph{ApJ}, \textbf{588},
  992--1002 (2003)

\bibitem[Tam et~al.(2002)]{trk02}
Tam, C., Roberts, M. S.~E., and Kaspi, V.~M., \emph{ApJ}, \textbf{572},
  202--208 (2002).

\bibitem[{Tam} and {Roberts}(2003)]{tr03}
{Tam}, C., and {Roberts}, M.~S.~E., \emph{ApJ}, \textbf{598}, L27--L30 (2003).

\bibitem[Torii et~al.(1999)]{ttd+99}
Torii, K., Tsunemi, H., Dotani, T., Mitsuda, K., Kawai, N., Kinugasa, K.,
  Saito, Y., and Shibata, S., \emph{ApJ}, \textbf{523}, L69--L72 (1999).

\bibitem[Kaspi et~al.(2001)]{krv+01}
Kaspi, V.~M., Roberts, M. S.~E., Vasisht, G., Gotthelf, E.~V., Pivovaroff, M.,
  and Kawai, N., \emph{ApJ}, \textbf{560}, 371--377 (2001).

\bibitem[Rots et~al.(1998)]{rjm+98}
Rots, A.~H., Jahoda, K., Macomb, D.~J., Kawai, N., Saito, Y., Kaspi, V.~M.,
  Lyne, A.~G., Manchester, R.~N., Backer, D.~C., Somer, A.~L., Marsden, D., and
  Rothschild, R.~E., \emph{ApJ}, \textbf{501}, 749--757 (1998).

\bibitem[Gotthelf et~al.(2000)]{gvbt00}
Gotthelf, E.~V., Vasisht, G., Boylan-Kolchin, M., and Torii, K., \emph{ApJ},
  \textbf{542}, L37--L40 (2000).

\bibitem[{Cusumano} et~al.(2001)]{cmm+01}
{Cusumano}, G., {Mineo}, T., {Massaro}, E., {Nicastro}, L., {Trussoni}, E.,
  {Massaglia}, S., {Hermsen}, W., and {Kuiper}, L., \emph{A\&A}, \textbf{375},
  397--404 (2001).

\bibitem[{Crawford} et~al.(2001)]{ckm+01}
{Crawford}, F., {Kaspi}, V.~M., {Manchester}, R.~N., {Lyne}, A.~G., {Camilo},
  F., and {D'Amico}, N., \emph{ApJ}, \textbf{553}, 367--374 (2001).

\bibitem[{Markwardt} et~al.(2003)]{mis03}
{Markwardt}, C.~B., {Ibrahim}, A.~I., and {Swank}, J.~H., \enquote{{XTE
  J1810-197},} in \emph{International Astronomical Union Circular}, 2003, pp.
  2--+.

\bibitem[{Ibrahim} et~al.(2003)]{ims+03}
{Ibrahim}, A.~I., {Markwardt}, C., {Swank}, J., {Ransom}, S., {Roberts}, M.,
  {Kaspi}, V., {Woods}, P., {Safi-Harb}, S., {Balman}, S., {Parke}, W.,
  {Kouveliotou}, C., {Hurley}, K., and {Cline}, T., \emph{ArXiv Astrophysics
  e-prints} (2003).

\bibitem[{Gotthelf} et~al.(2003)]{ghbb03}
{Gotthelf}, E.~V., {Halpern}, J.~P., {Buxton}, M., and {Bailyn}, C.,
  \emph{ArXiv Astrophysics e-prints} (2003).

\end{thebibliography}

\end{document}